\def\pdot {\dot P}
\def\ltsima{$\; \buildrel < \over \sim \;$}
\def\lsim{\lower.5ex\hbox{\ltsima}}
\def\gtsima{$\; \buildrel > \over \sim \;$}
\def\gsim{\lower.5ex\hbox{\gtsima}}

\def\uu {4U\,0142$+$614}
\def\oo {1E\,1048.1$-$5937}
\def\kes {1E\,1841$-$045}
\def\axj {AX\,J1844$-$0258}
\def\rx {1RXS\,J170849$-$400910}
\def\ee {1E\,2259$+$586}

\def \SAIT #1 #2 {{\em Mem.\ Soc.\ Astron.\ It.\/} {\bf #1}, #2}
\def \MESS #1 #2 {{\em The Messenger\/} {\bf #1}, #2}
\def \ASTRNACH #1 #2 {{\em Astron. Nach.\/} {\bf #1}, #2}
\def \AAP #1 #2 {{\em Astron. Astrophys.\/} {\bf #1}, #2}
\def \AAL #1 #2 {{\em Astron. Astrophys. Lett.\/} {\bf #1}, L#2}
\def \AAR #1 #2 {{\em Astron. Astrophys. Rev.\/} {\bf #1}, #2}
\def \AAS #1 #2 {{\em Astron. Astrophys. Suppl. Ser.\/} {\bf #1}, #2}
\def \AJ #1 #2 {{\em Astron. J.\/} {\bf #1}, #2}
\def \ANNREV #1 #2 {{\em Ann. Rev. Astron. Astrophys.\/} {\bf #1}, #2}
\def \APJ #1 #2 {{\em Astrophys. J.\/} {\bf #1}, #2}
\def \APJL #1 #2 {{\em Astrophys. J. Lett.\/} {\bf #1}, L#2}
\def \APJS #1 #2 {{\em Astrophys. J. Suppl.\/} {\bf #1}, #2}
\def \APSS #1 #2 {{\em Astrophys. Space Sci.\/} {\bf #1}, #2}
\def \ASR #1 #2 {{\em Adv. Space Res.\/} {\bf #1}, #2}
\def \BAIC #1 #2 {{\em Bull. Astron. Inst. Czechosl.\/} {\bf #1}, #2}
\def \JSQRT #1 #2 {{\em J. Quant. Spectrosc. Radiat. Transfer\/} {\bf #1}, #2}
\def \MN #1 #2 {{\em Mon. Not. R. Astr. Soc.\/} {\bf #1}, #2}
\def \MEM #1 #2 {{\em Mem. R. Astr. Soc.\/} {\bf #1}, #2}
\def \PLR #1 #2 {{\em Phys. Lett. Rev.\/} {\bf #1}, #2}
\def \PASJ #1 #2 {{\em Publ. Astron. Soc. Japan\/} {\bf #1}, #2}
\def \PASP #1 #2 {{\em Publ. Astr. Soc. Pacific\/} {\bf #1}, #2}
\def \NAT #1 #2 {{\em Nature\/} {\bf #1}, #2}

\documentstyle{memsait}
\input epsf.tex
\begin{opening}
\title{OBSERVATIONS OF ANOMALOUS X--RAY PULSARS} 
\author{GIANLUCA ISRAEL$^{1,2}$, SANDRO MEREGHETTI$^{3}$, LUIGI
STELLA$^{1,2}$}
\institute{$^1$Osservatorio Astronomico di Roma, Via Frascati 33,
Monteporzio Catone, Italy; stella and israel@oar.mporzio.astro.it .\\
$^2$Affiliated to I.C.R.A.\\
$^3$Istituto di Fisica Cosmica G.Occhialini, CNR, Via Bassini 15,
Milano, Italy; sandro@ifctr.mi.cnr.it .}
\date{} 
\end{opening}

\begin{document}

\oddpagefooter{}{}{} 
\evenpagefooter{}{}{} 
\
\bigskip

\begin{abstract}
We review recent results obtained from observations of Anomalous 
X--ray Pulsars at different wavelengths (X--rays, Optical, IR and Radio) 
with particular emphasis on results obtained by {\em Beppo}SAX. 
Proposed models for AXPs are briefly presented and discussed in 
the light of these results.  
\end{abstract}

\section{Introduction}
Coherent X--ray pulsations from \ee\ were
discovered at the end of seventies. Only few years ago it was
recognized however that a number of X--ray pulsators, including 
\ee, posses peculiar
properties which are very much at variance with those of known accreting
pulsars in X--ray binaries.
These objects, initially suggested as a homogeneous new class of pulsators
(Mereghetti \& Stella 1995; van Paradijs et al. 1995), have been
named in different ways, reflecting our ignorance of their nature:
Very Low Mass X--ray Pulsars, Braking Pulsars, 6\,s Pulsars,
Anomalous X--ray Pulsars.
The latter designation (AXPs) has became the most popular and will be used
hereafter.

The nature of AXPs is one of the most
challenging unsolved problems of Galactic high energy astrophysics. Over
the last few years there has been
a remarkable observational and theoretical effort to unveil their nature.
Although we can be reasonably confident that AXPs are magnetic rotating
neutron stars (NSs), their energy production mechanism is still
uncertain; it is also unclear whether they are solitary objects or are
in binary systems with very low mass companion.
As a consequence, different production mechanisms for the observed
X--ray emission have been proposed, involving either accretion or
the dissipation of magnetic energy.

\begin{table}[tbh]
\centerline{\bf Tab. 1 - Anomalous X--ray Pulsars (AXPs)}
\begin{center}
\vspace{-3mm}
\begin{tabular}{|l|c|c|c|c|}

\hline

{\bf  SOURCE}       &{\bf P (s)}  &{\bf $\pdot$  (s s$^{-1}$)}  &{\bf SNR}       &  {\bf    SPECTRUM  }  \\
                   &             &                    & {\bf d (kpc)/age (kyr)}  &  {\bf kT$_{BB}$(keV)/$\alpha_{ph}$}    \\
\hline
1E~1048.1--5937    & 6.45 &(1.5--4)$\times$10$^{-11}$ &    --                 & BB+PL  [3]  \\
                   & [1]  &      [2,3]                &    $\sim$5 / --                   & $\sim$0.64 / $\sim$2.5  \\
1E~2259+586        & 6.98 & $\sim$5$\times$10$^{-13}$ & G109.1--0.1 [7,8,9]   & BB+PL [9]      \\
                   & [4]  &       [5,6]               &   4--5.6 / 3--20      & $\sim$0.44  / $\sim$3.9 \\
4U~0142+614         & 8.69 & $\sim$2$\times$10$^{-12}$ &    --                 & BB+PL [11,12]        \\
                   & [10] &       [11]                &    $\sim$1--2 / --                   & $\sim$0.38  /  $\sim$3.9   \\
RXSJ170849--4009   &11.00 & 2$\times$10$^{-11}$       &    --                 & BB+PL [13,14]          \\
                   & [13] &       [15]                &    $\sim$10 / --                   & 0.45 / 2.6  \\
1E~1841--045       &11.77 & 4.1$\times$10$^{-11}$     &  Kes 73 [18,19]       & PL  [20]      \\
                   & [16] &     [17]                  &   6--7.5 / $\lsim$3   &  -- / $\sim$3.4   \\
{\em AX~J1844--0258}   & 6.97 & --                        & G29.6+0.1 [22]        & BB  [21]   \\
 {\em (candidate)}    & [21] &    &     $\leq$15 / $<$8      & $\sim$0.7 keV / --   \\
\hline
\end{tabular}
\end{center}

[1] Seward et al. 1986;
[2] Mereghetti 1995;
[3] Oosterbroek et al. 1998;
[4] Fahlman \& Gregory 1981;
[5] Baykal \& Swank 1996;
[6] Kaspi et al. 1999;
[7] Hughes et al. 1984;
[8] Rho \& Petre 1997;
[9] Parmar et al. 1998;
[10] Israel et al. 1994;
[11] Israel et al. 1999a;
[12] White et al. 1996;
[13] Sugizaki et al. 1997;
[14] Israel et al. 2001a,
[15] Israel et al. 1999b;
[16] Vasisht \& Gotthelf 1997;
[17] Gotthelf et al. 1999;
[18] Sanbonmatsu \& Helfand 1992;
[19] Helfand et al. 1994;
[20] Gotthelf \& Vasisht 1997;
[21] Torii et al. 1998;
\end{table}

The properties that distinguish AXPs from known magnetic ($\geq$
10$^{12}$ G) accreting X--ray pulsars found in
High and Low Mass X--Ray Binaries (HMXBs and LMXBs) are the following:
\begin{itemize}
\item spin periods in a narrow range ($\sim$6--12\,s)
compared with the much broader distribution (0.069 -- $\sim$10$^4$\,s)
observed in HMXRB pulsars; \vspace{-3mm}
\item no conspicuous optical counterparts (see Section\,6),
with upper limits which rule out the presence of massive companions,
like OB (super)giants or Be stars;\vspace{-3mm}
\item very soft and absorbed X--ray spectra: a black body with characteristic
temperature in the 0.4--0.6\,keV range and a steep power--law with photon
index in the 2.5--4 interval (suggesting that perhaps a large part of 
the total luminosity is hidden in the EUV band);\vspace{-3mm}
\item relatively low X--ray luminosity  ($\sim10^{34}$--10$^{36}$ erg
s$^{-1}$) compared with that of HMXB pulsars;\vspace{-3mm}
\item relatively low flux variability on timescales from hours to
years;\vspace{-3mm}
\item relatively stable spin period evolution, with long term spin--down
trend (see Section\,5);\vspace{-3mm}
\item a very flat distribution in the Galactic plane and three clear 
associations with supernova remnants
(SNRs; suggesting a young population).
\end{itemize}

There are currently (February 2001) five ascertained members of the AXP
class (see Table 1) plus one likely candidate.
This review, after a brief presentation of theoretical models 
(Section\,2), concentrates on the recent results inferred from observations
of AXPs at different wavelengths (X--rays, Section\,4 and 5;
optical and IR, Section\,6; radio, Section\,7). The implications of these
results for the proposed theoretical models are briefly discussed.

\section{Theoretical Models}
Theoretical models for AXPs can be classified into two main classes
depending on the mechanism that is supposed to power their X--ray emission:
accretion or magnetic field decay. The former class includes both isolated
NSs and NSs in binary systems.

Accretion from matter in a overdense region was first proposed
for \uu\ based on its apparent association with a molecular cloud.
However this would imply that AXPs move at low velocities in the
ambient medium (Israel et al. 1994).  Mereghetti \& Stella (1995) 
proposed that
AXPs form an homogeneous subclass of accreting neutron stars, perhaps
members  of very low mass X--ray binaries (VLMXBs), which are
characterized by lower luminosities and higher magnetic fields
($B$$\sim$10$^{11}$\,G) than accreting neutron  stars in classical
LMXBs.
On the other  hand, Van Paradijs et al. (1995) proposed that AXPs are the
result of a common envelope and spiral--in evolution of a neutron star
and its massive companion. This ends up in the complete disruption 
of the companion star
after the so--called Thorne--Z\.ytkov stage. Based on the AXPs--SNRs
association, Chatterjee et al. (2000) and Perna et al.
(2000), proposed instead that AXPs accrete from a fossil disk 
made of matter falling
back onto the neutron star after its birth.  Alpar (2000) proposed
that AXPs are NSs accreting from the debris of their SNRs, in a phase
following the SGR stage, or, alternatively, NSs accreting from a companion
star in a rare path of LMXB evolution. In these scenarios the P/\.P 
does not provide a reliable good estimator of the pulsar age; 
rather it should represent
an asymptotic value approached by the pulsar. The equilibrium period should
depend on the trapped angular momentum in the residual accreting
matter (Alpar 2000). In the accretion scenario the narrow period
distribution of AXPs can be accounted for by either limiting the magnetic
field of the NS and strength of the propeller wind emission (Marsden
et al. 1999) or using ADAF models and an appropriate distribution of
magnetic field, initial spin and accretion disk mass (Chatterjee \&
Hernquist 2000).

Finally, Thompson  \& Duncan (1993, 1996)
proposed that  AXPs are ``magnetars'', isolated  neutron stars with a
super-strong magnetic field ($\sim10^{14-16}$ G; for a review see also
Thompson, this book). Magnetars were
originally  proposed to explain the properties of soft  $\gamma$--ray
repeaters. These were later determined to exhibit pulse periods
(8.05\,s, 7.5\,s  and 5.16\,s in SGR\,0526--66, SGR\,1806--20 and 
SGR\,1900+14, respectively) and period derivatives similar to those of
AXPs (Kouveliotou et al. 1998;
Hurley et al. 1999). If this connection proved correct, AXPs might
be some sort of quiescent analogous of soft $\gamma$--ray repeaters.
In the framework of the magnetar model, Colpi et al. (2000)
showed that it is possible to account for the narrow period distribution
of AXPs if the initial magnetic field (in the 10$^{15}$--10$^{16}$ G range)
of magnetars decays significantly on a timescale of the order of
10$^{4}$ years.

\section{The AXP sample}
Table\,1 lists the main characteristics of the five X--ray pulsars which
form the AXP class. The new AXP candidate, \axj, is also included.

Recently a possible connection of AXPs with Soft $\gamma$--Repeaters (SGRs,
for a review see Kouveliotou, this book) has been proposed
(Kouveliotou et al. 1998, 1999; Hurley et al. 1999); this builds on several
observational similarities, such as the range of spin periods, their
derivatives, and the possible association with SNRs. In the  case of SGRs large
offsets from the SNR centers were measured which might imply that AXPs may
eventually evolve into SGRs as they age and move away from the SNR centre
(Gaensler 2000). This would require a difference of 10--100\,kyr in the age
of AXPs and SGRs; therefore it would be difficult to explain, in the light
of the inferred period derivatives, the similarity of periods.
We also note that a 226\,ms radio pulsar has been recently discovered in
the SNR previously associated with the Soft $\gamma$--Repeater SGR\,1900+14
(Lorimer \& Xilouris 2000). Although it is unclear which of the two objects
(if any) is associated to the SNR, a recent reanalysis of 
the distance indicators (Case \& Bhattacharya 1998) places the SNR at
10$\pm$3\,kpc, consistent with the position of the newly discovered
radio pulsar.
The alternative possibility is that SGRs are born with a different velocity
distribution than AXPs and, therefore, the two classes cannot be drawn from
the same parent population. The AXPs--SGRs connection is still an open issue.

\section
{The {\em Beppo}SAX view}
Observations by different X--ray satellites
have often been used to check for the presence of spectral and flux variations
in AXPs. These, however, are affected by the uncertainties introduced by
comparing the results of different instruments covering different energy
ranges, etc. The  {\em Beppo}SAX satellite, with a coherent set
of instruments covering the 0.1--10\,keV range with spectral resolution of 3--10,
observed the whole sample of AXPs.
Spectral results from  {\em Beppo}SAX observations have already been
reported for three AXPs, \ee\ (Parmar et al. 1998), \oo\ (Oosterbroek et al.
1998), and \uu\ (Israel et al. 1999).
In the following we will report on the spectral and timing properties of another
AXPs, \rx\ (Israel et al. 2001a), observed by  {\em Beppo}SAX
in March--April 1999.
The  {\em Beppo}SAX observation of 1E1841--045
is not included in this report due to strong contamination
from the SNR Kes\,73 (timing results have been reported by Gotthelf et
al. 1999). The candidate AXP \axj\ has not yet been observed by {\em Beppo}SAX.

\begin{figure}
\epsfysize=7.9cm 
\hspace{1.5cm}\epsfbox{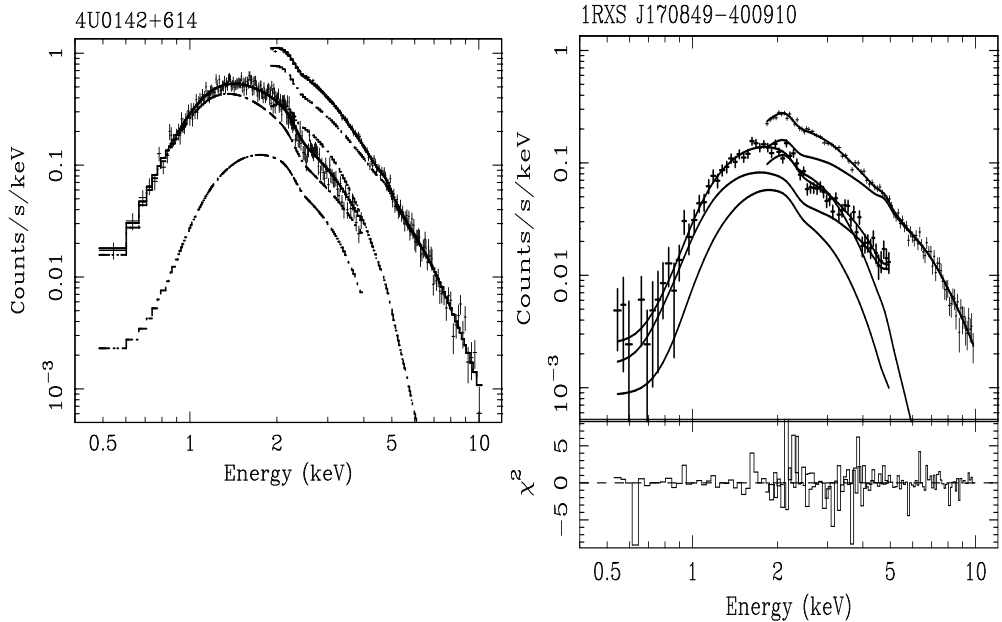}
\caption[h]{LECS and MECS energy spectra of two AXPs, \uu\ (left; adapted
from Israel et al. 1999) and
\rx\ (right). For the latter source the residuals (in units of $\chi^2$)
of the best fit are also shown. The power--law and blackbody components
are also shown.}
\end{figure}

The relatively bright source \rx\ was discovered by ROSAT at the beginning
of its mission; however only in 1997 this source attracted much attention 
because of the discovery of $\sim$11\,s pulsations with   ASCA
(Sugizaki et al. 1997). Based on the pulse period and unusually soft
X--ray spectrum the source was tentatively classified as a candidate AXP.
This interpretation was confirmed thanks to ROSAT HRI
observations which provided the first measurement of the period
derivative \.P$\sim$2$\times$10$^{-11}$\,s\,s$^{-1}$ and a more accurate
X--ray position from which it was possible, through optical imaging, to
confidently exclude the presence of a massive companion star
(Israel et al. 1999b). \rx\ has been monitored with the
{\em Rossi}XTE since 1998;  a sudden spin--up event,
suggestive of a ``glitch'' from a highly magnetised NS,
 was recorded  in September 1999
(Kaspi et al. 2000a; see also Section\,5).

\begin{figure}
\epsfysize=9cm 
\hspace{4.5cm}\epsfbox{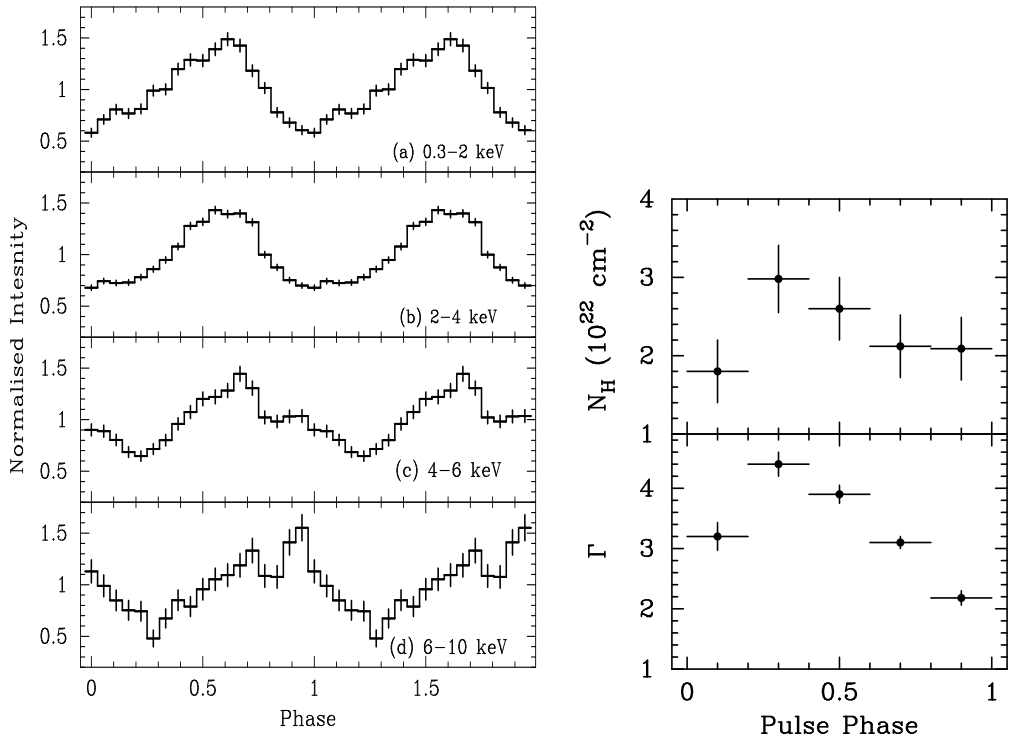}
\vspace{1.cm}
\caption[h]{\rx\ MECS and LECS light curves folded to the best period
(P=10.99915\,s) for four different energy intervals (left panels).
For clarity two pulse cycles are shown. Zero phase was (arbitrarily)
chosen to correspond to the minimum in the 0.3--2\,keV folded light curve.
The results of the pulse phase spectroscopy are also reported for the two
free spectral parameters (right panels).}
\end{figure}

{\em Beppo}SAX observed this source on 31 March -- 1 April 1999 with
the Narrow field Instruments: the Low--Energy Concentrator Spectrometer
(LECS; 0.1--10~keV; Parmar et al. 1997; 26\,ks effective exposure time)
and the Medium--Energy Concentrator Spectrometer (MECS; 1.3--10~keV;
Boella et al. 1997; 52\,ks effective exposure time).
A simple power--law model (as well as any other single component model)
did not fit well the data ($\chi^2$ of 1.42 for 199 degrees of freedom).
A better fit (see right panel of Figure 1) was obtained including a soft
thermal component in analogy with similar results obtained for other AXPs
($\chi^2$ of 1.12 for 197 degrees of freedom). 
The best fit was obtained for an absorbed,
$N_H$=(1.46$\pm$0.02)$\times$10$^{22}$\,cm$^{-2}$, power--law with
photon index 2.6$\pm$0.2 and a blackbody component with a characteristic
temperature of 0.45$\pm$0.03\,keV (90\% c.l. reported; Israel et al. 2001a).
The unabsorbed  1--10\,keV
 flux was 6$\times$10$^{-11}$\,erg\,s$^{-1}$. The blackbody
component accounts for about 30\% of the total observed flux. Figure\,1
compares the spectral shape and components of \rx\ and \uu\ as seen by
{\em Beppo}SAX. \rx\ is therefore the fourth AXP for which a two
componet spectrum (steep power--law plus soft blackbody) has been detected.
\begin{figure}
\epsfysize=8cm 
\hspace{1.5cm}\epsfbox{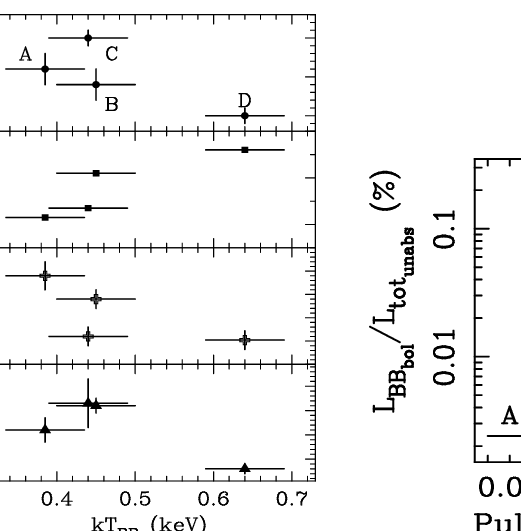}
\caption[h]{The main spectral properties of AXPs (A=\uu, B=\ee, C=\rx\ and 
D=\oo) observed by
{\em Beppo}SAX are reported as a function of the characteristic blackbody
temperature (left panels). Error bars in the blackbody radius reflects
also the uncertainty in the assumed distances. Absorption column values
are in units of 10$^{21}$\,cm$^{-2}$. The correlation between
AXP pulsed fractions and the bolometric $L_{\rm BB_{\rm bol}}$ and
the total unabsorbed 0.1--10\,keV luminosity $L_{\rm tot_{\rm unabs}}$ 
ratio is also shown (right panel).}
\end{figure}

A further interesting result of the \rx\ {\em Beppo}SAX observation was
the detection of a clear pulse shape change as a function of energy
(see left panels of Figure\,2).
In particular the minimum in the lowest energy interval (0.1--2\,keV)
corresponds to a maximum in the 6--10\,keV band.
The pulsed fraction (semiamplitude of
modulation divided by the mean source count rate) decreases from
40\% to 30\% from the lowest to the highest energy band, respectively.
Pulse phase spectroscopy was carried out with the MECS data
(due to poor statistics at low energies the data from the LECS data 
were not used). By keeping the parameters of the blackbody component 
fixed at the phase--averaged value, a $\sim$3$\sigma$ significant variation 
in the power--law photon index was found. The large 
uncertainties prevent the detection of possible variations in $N_H$ 
(see right panels of Figure\,2). Although small energy--dependent changes 
in the pulse shape were already suggested in the past for
\uu\ (Israel et al. 1999a; Paul et al. 2000), the variations of \rx\ are 
highly significant and likely arising from a changing power--law slope. 
The lack of any conspicuous change in the pulse profiles of AXPs 
was used in the past to argue against the
possibility that these sources are accreting X--ray objects. More sensitive
studies will yield additional information on the spectral changes
causing the pulse shape variations, extend the energy range of the
source detection above 10\,keV, and allow to look for cyclotron signatures.

The spectral properties of \rx\ reported above are plotted in Figure\,3
(left panels) together with the {\em Beppo}SAX results obtained
for the other three AXPs for which good spectral data are
available.
It is apparent that the spectral properties of AXPs are relatively 
homogeneous. \oo\ has the highest blackbody temperature and  largest 
(absorbed) $L_{BB}$/$L_{tot_{\rm abs}}$ ratio.
Note that the $L_{BB}$/$L_{tot_{\rm abs}}$ ratio is not strongly dependent
on the power--law photon index. The AXPs  with the largest (\rx) and 
smallest (\oo) absorption have nearly the same photon index, suggesting 
that the detection of a steep power--law
in AXPs is not an artifact due to a large $N_H$ value.

The picture
becomes quite different when extrapolating the total unabsorbed fluxes in the
0.1--10\,keV range and considering the bolometric blackbody flux; this gives
in all cases   a $L_{BB_{\rm bol}}$/$L_{tot_{\rm unabs}}$ ratio 
smaller than 30\%. Despite the $\sim$10$^3$ ratio 
in total unabsorbed flux between the brightest (\uu) and faintest (\rx)
AXPs, the blackbody component varies by only a factor of $\sim$10,
strongly suggesting that the energy budget of AXPs is largely dominated by
the power--law. Another interesting result is shown in the right panel
of Figure\,3 where the 0.1--10\,keV pulsed fractions 
are plotted as a function of $L_{BB_{\rm bol}}$/$L_{tot_{\rm unabs}}$. 
This correlation strongly suggests that the thermal component is mainly 
responsible for the pulsations component. This is also in agreement with 
the fact that \uu, the brightest AXP, has also the lowest known pulsed
fraction (Israel et al. 2001b).

\section{Further X-ray observations}

Several results from  X--ray observations of AXPs and related objects
were reported in the last year. In particular the pulsation stability
of AXPs was compared to that of conventional accreting X--ray pulsars
and isolated radio pulsars. \rx\ and \ee\ were the
first two AXPs for which a systematic study was carried out and
phase--coherent timing solutions obtained by means of {\em Rossi}XTE data
(Kaspi et al. 1999). These two sources were found to be quite stable
rotators with phase residuals of only $\sim$1\%, comparable to or smaller
than those measured for most radio pulsars.
However, in September 1999 the {\em Rossi}XTE satellite detected a sudden
spin--up event in \rx\  which was originally interpreted as a ``glitch''
similar to that observed in the Vela and other young radio pulsars
(Kaspi et al. 2000a). However, we note that in principle glitches could 
also be detected in accreting spinning--down X--ray sources 
(with a sufficiently high magnetic field strength) if they are in a low
noise level phase, as indeed AXPs are known to be.
A way to distinguish, in the near future, whether \rx\ experienced 
a radio pulsar--like glitch or, perhaps, an accreting X--ray pulsar--like
spin--up behaviour would be to accurately monitor the period history 
after the event. Unfortunately these results have not been reported so far.

Also the pulsations stability of \oo\ was studied in great detail with
ASCA (Paul et al. 2000) and  {\em Rossi}XTE (Kaspi et al. 2000b) data.
In the latter case the sampling proved insufficient to find a phase--coherent
solution due to a high noise level (changing spin--down rate).
The observed deviations from simple spin--down were
found to be inconsistent with a single glitch event. Such period changes do
not seem to be accompanied by pulse shape or spectral changes (in ASCA) or
even flux variations as observed in accreting X--ray pulsars (Paul et al. 2000;
Kaspi et al. 2000b). In this respect \oo\ was suggested to be a transition
object between SGRs and AXPs (Kaspi et al. 2000b). Based on   earlier
{\em Rossi}XTE data, Baykal et al. (2000a) found that the level of pulse
fluctuations in \oo\ is consistent with the typical noise level of accretion
powered pulsars (Baykal \& \"Ogelman 1993; Bildsten et al. 1997).
Also the results of further {\em Rossi}XTE observations of \ee\
suggest that the source is perhaps in a fairly stable accretion phase 
with a constant X--ray luminosity and spin--down rate (Baykal et al. 2000a).
An interesting result has  been recently reported for the high--mass
accreting  X--ray pulsar 4U\,1907+09. This is  the only known X--ray
pulsar (besides AXPs) which has always been spinning--down since the discovery
of pulsations at 440\,s in its X--ray flux in 1983. This result, based 
again on {\em Rossi}XTE data, shows that 4U\,1907+09 during 1996--1998 
has been a rather stable 
rotator with a spin--down rate only a factor four greater than that of \ee\
(Baykal et al. 2000b). Furthermore the long--term noise strength of
4U\,1907+09 is one order of magnitude lower than that of \ee;
the conclusion is that the existence of an accreting source displaying
persistent spin--down shows that quiet spin--down trends do not necessarily
imply that the sources are not accreting.

No new results have been reported concerning the search of
orbital signatures by means of delays in the pulse arrival times of AXPs.
In fact after the tight upper limits inferred on the possible
companion star mass of \oo\ and \ee\ (Mereghetti et al. 1998),
and \uu\ (Wilson et al. 1999) based on {\em Rossi}XTE data, none of the
other three AXPs has been studied in this respect. The lack of detectable
Doppler effect has been used by many authors to favor the magnetar
scenarios. It is worth mentioning that the pulse arrival time
delays remain undetected also in 4U\,1626--674, an X--ray pulsar with a
very low--mass companion star (likely a light He--burning or He white 
dwarf; see Chakrabarty 1998) in a 42\,min orbital period binary. 
If AXPs were in binary systems with companion stars
similar to that of 4U\,1626--674, expected delays would be below the
current upper limits.

Finally Marsden \& White (2001) recently found a correlation between 
spin--down rates of AXPs and SGRs and the L$_{\rm BB_{bol}}$/L$_{\rm PL}$ 
ratio concluding that, regardless the nature of these sources, they are likely
objects with similar emission mechanisms.

\section{Optical/IR follow--up observations}
Similar to other Galactic X--ray sources, the identification of AXPs at
optical and IR wavelengths suffers from the comparatively poor
spatial resolution of X--ray telescopes which is often inadequate to cope with
the crowded Galactic plane fields in which AXPs lie, and the strong extinction
in their direction.
Although the X--ray positional accuracy  is not yet
sufficient to unambiguously identify the possible optical/IR counterparts
of AXPs, it is now possible to sort out a number of good candidates or,
at least, put tight constraints on the optical/IR--to--X--ray energy distribution
of these sources.

This is the case of the recent proposed optical counterpart of \uu,
a faint relatively blue object ($R$=$25$; $V-R$=$0.63$) which, in the
color--magnitude diagram, lies half--way between the main sequence 
and the track of 0.6\,M$_{\odot}$ white dwarfs (Hulleman et al. 2000a; Keck
observations). However the optical counterpart candidate position 
of \uu\ (within the Einstein HRI error circle; White et al. 1987) is outside
the EXOSAT LEIT error circle (White et al. 1987) and at the edge of the 
ROSAT PSPC one. A more accurate X--ray position is clearly needed in order 
to confirm this result.
Since none of the accretion--based scenarios considered by Hulleman et al. 
(2000a) fit the optical and X--ray data these authors conclude
that the measurements may be in agreement with a magnetar. However 
no detailed models have yet been developed for the optical emission of
magnetars (Hulleman et al. 2000a). We note that the
optical data are also in agreement with at least two other scenarios: 
(a) a binary system hosting a white dwarf companion, which accounts
for the optical emission, and an accreting neutron star producing the X--ray
emission, (b) a disk around an isolated NS with a more realistic choice 
of values for the disk size, illumination and inclination 
(Israel et al. 2001c).

Hulleman et al. (2000b) carried out also a relatively deep search for
the optical counterpart  of \ee\ based on Keck
observations. No object was found down to  limiting magnitudes of
$R$=$25.7$ and $I$=$24.3$. These authors conclude that
it is unlikely that \ee\ is an isolated NS accreting from a residual disk. 
Also in this case, however a binary system with a helium--burning 
0.3\,M$_{\odot}$ or a white dwarf companion star, or a magnetar scenario 
cannot be excluded yet.

Optical observations of the field of \rx\ were obtained 
by Israel et al. (1999b). These authors found  that the possible counterpart
cannot be a massive early type star (a distant and/or absorbed OB star
would appear more reddened). However the images were taken from a
1.5\,m telescope and are not deep enough to constrain any other proposed
theoretical scenario such as a low mass companion, a residual disk or a
magnetar.

\begin{table}[tb]
\centerline{\bf Tab. 2 - Current Optical/IR upper limits and measurements}
\begin{center}
\vspace{-3mm}
\begin{tabular}{|l|c|c|c|c|c|c|l|}
\hline
SOURCE & $B$ & $V$ & $R$ & $I$ & $J$ & $K$ & Ref. \\
\hline
1E~1048.1--5937    &  $>$24.5 & $>$24.5 & $>$24.3 & ---     & ---     & --- 
& [1]\\
1E~2259+586        &  $>$25   & $>$24   & $>$25.7 & $>$24.3 & $>$19.6 & $>$18.4& [2,3]\\
4U~0142+614        &  ---     & 25.6    &   25    &   25    & $>$19.6 & $>$16.9& [4,3]\\
RXSJ170849--4009   &  $>$20   & $>$25   & $>$26.4 & $>$25   & ---     & --- 
& [5,6]\\
1E~1841--045       &  $>$23   & $>$22   & $>$24   & ---     & $>$19   & $>$17   &[7] \\
\hline
\end{tabular}
\end{center}
Note:  Values are taken from  [1] Mereghetti et al. 1992, 
[2] Hulleman et al. 2000b, [3] Coe \& Pightling 1998,
[4] Hulleman et al. 2000a, [5] Israel et al. 1999b, [6] Israel et al.
2001a, [7] Mereghetti et al. 2001.
\end{table}

Recently Mereghetti et al. (2001) reported on the first search for
optical/IR counterpart in the field of \kes\ located at the center of
the SNR Kes\,73 (mainly with the 1.5\,m and 3.5\,m class ESO telecopes).
Similar to the previous case, the results are not very
constraining, due to the high extinction in the direction of the source 
($A_V$$\geq$11).
No detailed reports on optical/IR  observations of \axj\ have yet been
published.
Table\,2 summarises the current optical/IR upper limits and measurements
of AXPs.

\section{Radio follow--up}
If AXPs are isolated NSs that emit radiation in a fashion similar to 
rotation powered pulsars then they might shine and pulsate also in the 
radio band. We observed the fields of four southern
AXPs, namely \oo, \kes, \axj\ and \rx, with the Parkes Observatory and
typical exposures of 10--20\,ks for each source. The sampling time
was in the 1--1.2\,ms range with a beam aperture of $\sim$14' (at 1.4\,GHz).
Sampled duty cycles were in the $\sim$0.001--20\% range, while 
dispersion measures (DM) up to 10 times larger than the
Galactic values in the direction of observed AXPs were sampled 
in order to take into account also the possible presence of local matter. 
Setting the detection threshold at a signal to noise ratio of 10, an average 
value of $\sim$70$\mu$Jy at 1.4\,GHz can be reasonably assumed for the 
upper limit to the radio flux of the sample of observed AXPs observed. 
The search for coherent signals at the X--ray period (we used the 
$P$ and \.P values reported in Table\,1
to infer the period interval for the search) did not detect any significant
signal (Burderi et al. 2001).

\section{The Future}
After more than 20 years from the discovery of pulsations from \ee\ 
the nature of AXPs is still uncertain.
Important limiting factors in the study of AXPs are: (i) the relatively
low X--ray positional accuracy provided so far by X--ray telescopes 
together with the difficulties
in obtaining deep optical/IR images (needed to sample faint objects)
from the largest ground--based telescopes, (ii) the lack of spectral
information above 10\,keV where many cosmic X--ray
sources (X--ray pulsars especially) display important features,  (iii) 
the paucity of unambiguous predictions by theoretical models, and (iv) the
small numbers in the currently known AXP population.
The present generation of X--ray astronomy satellites
(Chandra and {\em Newton}XMM) and large ground--based telescopes (VLT, Gemini,
Subaru and KeckII) opens new prospectives in the field.
The unrivalled spatial resolution  and absolute source positioning accuracy
($\leq$1'') of Chandra will provide the best chance to unambiguously 
identify the counterparts of AXPs.
Moreover assessing that the AXP candidate \axj\ (which has shown a 
factor of variability larger than 10) is indeed an AXP would imply that 
the number of AXPs and their formation rate have been largely
underestimated. The high throughput of {\em Newton}XMM
and the relatively wide field of view of its instruments will presumably
allow to find other AXPs even at fainter fluxes, thus providing new 
important informations to understand the puzzling nature of AXPs.

\acknowledgements
The results reported in this review have been obtained with the contribution
of a number of scientists, who are part of large international collaboration
aimed at studying the AXPs. It is a pleasure to
thank all the people who have contributed to this
project: L. Angelini, L. Burderi, S. Campana, S. Covino, N. D'Amico,
F. Haberl, G. Marconi, R. Mignani, I. Negueruela, T. Oosterbroek,
A.N. Parmar, A. Possenti, and  N.E. White.

\end{document}